\documentclass[prd,twocolumn,showpacs,preprintnumbers,superscriptaddress,floatfix,nofootinbib]{revtex4-1}
\usepackage{multirow}
\usepackage{amsfonts}
\usepackage{graphicx,,booktabs}
\usepackage{amssymb}
\usepackage{amsmath,txfonts,ulem}
\usepackage{graphicx}
\usepackage{dcolumn}
\usepackage{color}
\usepackage{bm}
\usepackage{ulem}
\usepackage[colorlinks,
            citecolor=blue,
            anchorcolor=green,
            menucolor=orange,
            linkcolor=red,
            filecolor=red,
            runcolor=pink,
            urlcolor=blue,
            frenchlinks=red]{hyperref}

\allowdisplaybreaks[4]

\begin{document}

\title{\boldmath Systematic investigation of trace anomaly contribution in nucleon mass}

\author{Xiao-Yun Wang}
\email{xywang@lut.edu.cn}
\affiliation{Department of physics, Lanzhou University of Technology,
Lanzhou 730050, China}
\affiliation{Lanzhou Center for Theoretical Physics, Key Laboratory of Theoretical Physics of Gansu Province, and Key Laboratory of Quantum Theory and Applications of MoE, Lanzhou University, Lanzhou, Gansu 730000, China}

\author{Jingxuan Bu}
\email{jingxuanbu@yeah.net}
\affiliation{Department of physics, Lanzhou University of Technology,
Lanzhou 730050, China}

\date{\today}
\begin{abstract}
In this work, under the framework of vector meson dominance model, the trace anomaly contribution value inside neutrons are extracted for the first time based on vector meson photoproduction data. Furthermore, we systematically compare and analyze the trace anomaly contributions of protons and neutrons. The results show that the trace anomaly contributions of protons and neutrons are close, which indirectly confirms that their internal structures and dynamic properties may have certain similarities. In addition, the main factors affecting the extraction of the trace anomaly contribution of nucleons are discussed in detail. This study not only provides a theoretical basis for us to better understand the source of nucleon mass, but also makes a useful exploration and discussion on how to extract the trace anomaly contribution of nucleon more accurately in the future.

\end{abstract}


\maketitle

\section{introduction}

As the elementary particle of most visible matter, the study on the internal properties of nucleon is a valuable subject of quantum chromodynamics (QCD), where the source of the nucleon mass has always been a mystery. The Higgs mechanism explains only 2\% of it, most of the rest comes from the complex strong interactions. However, due to the weak gravitational interactions acting on the single nucleon, the source of the nucleon mass cannot be directly measured experimentally. Nevertheless, Ji et al. \cite{ji94,ji95,lorce} found a way to represent the nucleon mass in terms of quark and gluon energy theoretically, the nucleon mass is divided into four parts under their separation.

The separation starts from the QCD energy-momentum tensor (EMT) ${T}^{\mu \nu}$, which can be decomposed into the trace and traceless parts \cite{ji95}
\begin{align}\label{eq:EMT}
{T}^{\mu \nu}=\hat{T}^{\mu \nu}+\bar{T}^{\mu \nu}
\end{align}
Further, the trace parts is a sum of the quark mass and trace anomaly contributions, the traceless part is consist of the quark and gluon energy contributions, respectively.
\begin{equation}\label{eq:dEMT}
\begin{aligned}
&\hat{T}^{\mu \nu}=\hat{T}^{\mu \nu}_{a}(\mu^2)+\hat{T}^{\mu \nu}_{m}(\mu^2)
\\
&\bar{T}^{\mu \nu}=\bar{T}^{\mu \nu}_{q}(\mu^2)+\bar{T}^{\mu \nu}_{g}(\mu^2)
\end{aligned}
\end{equation}
And the tensor defines the QCD Hamiltonian operator as \cite{ji95}
\begin{equation}
H_{\mathrm{QCD}}=\int d^3 \vec{x}\ T^{00}(0, \vec{x})
\end{equation}

According to the above definition and decomposition of EMT, the corresponding four parts separations of the QCD Hamiltonian operator are obtained as \cite{ji95}
\begin{align}\label{eq:Hsum}
H_{QCD}=H_{q}+H_{g}+H_{m}+H_{a}
\end{align}
In  Reference \cite{ji95}, hadron mass is defined as the expectation value of the Hamiltonian operator in the rest frame,
\begin{align}\label{eq:Mn}
M=\frac{\left< P|H_{QCD}|P \right>}{\left< P|P \right>}\Bigg|_{\rm rest\ frame}
\end{align}
Thus, based on Eq.  \ref{eq:Hsum} and \ref{eq:Mn} the hadron mass decomposition is derived as \cite{ji95}
\begin{equation}\label{eq:Mnfour}
\begin{aligned}
&M_{q}=\frac{3}{4}\ (a-\frac{b}{1+\gamma_{m}})M
\\
&M_{g}=\frac{3}{4}\ (1-a)M
\\
&M_{m}=\frac{4+\gamma_{m}}{4(1+\gamma_{m})}bM
\\
&M_{a}=\frac{1}{4}(1-b)M
\end{aligned}
\end{equation}

Noted from the above, the specific value of the mass decomposition are determined by parameters $a$ and $b$, which is the fraction of the nucleon momentum carried by quarks and the trace anomaly parameter, respectively, $\gamma_{m}$ represents the anomalous dimensions \cite{ji94,ji95}. Notice that there has a new source of mass that appears in the last term, which is the so-called trace anomaly contribution, it depends only on the parameter $b$. Work \cite{Ji:2021pys} argues that the trace anomaly comes from the scale symmetry breaking among the regulation of the UV divergences, with the scheme independent. It contributes the nucleon mass by a Higgs-like mechanism and sets the scale of the other part of the mass separations. Besides, one can carry out the related mechanism of the quark confinement by studying the anomalous energy.

At present, there is some research on the proton trace anomaly contribution, such as lattice QCD calculation \cite{latice} and  the holographic calculation \cite{Hatta:2018sqd,Hatta:2019lxo,Hatta:2018ina}. However, there are discrepancies in the conclusions between the current works. For instance, in the latest lattice QCD calculation of the proton anomaly contribution, the result is 23\% \cite{latice}. The predictive ability of the holographic model is presented in the work \cite{nature} by JLab. They compared the predicted cross section of the holographic model with the experimental data, results show that the predicted cross section corresponding to the maximum and minimum trace anomaly contributions are very close, which means that the model can't determine the anomaly contribution at present.
 In addition, the trace anomaly contribution can also be calculated from the scattering between the quarkonium and the nucleon \cite{I,light,vmd,nature,Hatta:2018sqd,Hatta:2019lxo,Hatta:2018ina,wr}. This implies that we can study the trace anomaly based on the existing photoproduction experimental data by combining the above process with the vector meson photoproduction process under the vector meson dominance (VMD) model \cite{vmd}. In Reference \cite{nature}, the proton trace anomaly contribution is also calculated based on the $J/\psi$ photoproduction experimental data under the VMD model, but the results show that the anomaly contribution increases obviously with energy. That is, the result calculated at higher energy is larger, which can reach more than 20\%; the results extracted near the threshold are very small, only a few percent. In addition, there is also a work about the proton mass decomposition, which starts from the QCD EMT, and introducing the independent operators by Metz et al. \cite{AM}, they concluded that the anomaly contribution of proton is equal to zero. These inconsistent results indicate the necessity of the research on the nucleon trace anomaly contribution. In our previous work \cite{I}, we modified the VMD model to solve the energy dependence problem, and the error of the obtained results has been greatly reduced.
 
 Although there have been some studies on the trace anomaly contribution of proton \cite{I,light,nature,wr,latice}, not yet on neutron. One of the purposes of this work is to first extract the trace anomaly contribution of neutron mass under the VMD model. As another purpose, we will further analyze the factors affecting the nucleon anomaly contribution. After work \cite{I}, we further extended the work to other vector meson photoproduction processes with the improved method \cite{light}, showing that the trace anomaly contributions from different processes are not the same, although they are all relatively small. 
Such results bring the nucleon anomaly contribution another non-negligible error, these inspire us to further analyze the affecting factors in order to give a more accurate proportion of the trace anomaly part. A comprehensive analysis will be discussed in detail in summary, and the research direction of obtaining higher precision nucleon trace anomaly contribution is pointed out. In part before the summary, the influence of one of the parameter $\alpha_{s}$, which has a sizable inaccurate precision \cite{QCD 2016,2018,1992,effective 2007,2008,charge 2022,low 2022,HERMES 1997,HERMES 1998,g1(p) 1998,HERMES 2003,HERMES 2007,Novel 2022}, on the nucleon anomaly contribution is first studied numerically. To achieve this, we adopt two sets of $\alpha_{s}$ on the extraction named $\alpha_{A}$ and $\alpha_{B}$, one set of $\alpha_{s}$ ($\alpha_{A}$) comes from some previous prediction and fitting result \cite{phi 0.770,OR alpha s}, another group ($\alpha_{B}$) comes from a recent work \cite{machine} on predicting $\alpha_{s}$ based on machine learning methods and large amounts of experimental data.

The structure of the paper is as follows. After the introduction, the review of the calculation method VMD model is in Sec. \ref{sec:model}, the corresponding numerical results are in Sec. \ref{sec:result}, and finally Sec. \ref{sec:sum} gives a summary.

\section{FORMALISM}\label{sec:model}

In the VMD model, process $\gamma N\to VMN$ is related with process $VMN\to VMN$ and the forward differential cross
section of $\gamma N\to VMN$ reaction is expressed as \cite{vmd}
\begin{align}\label{eq:b vmd}
\frac{d\sigma_{\gamma N\to VMN}}{dt}\Bigg|_{t=t_{min}}=\frac{3\Gamma_{e^+ e^-} }{\alpha m_{V}}\left(\frac{p_{VMN}}{p_{\gamma N}}\right)^{2}\frac{d\sigma_{VMN\to VMN}}{dt}\Bigg|_{t=t_{min}}
\end{align}
where $\Gamma_{e^+ e^-}$ is the radiative decay width, the value of the electromagnetic coupling constant $\alpha$ is equal to 1/137 and $m_{V}$ is the vector meson mass. ${p}_{ab}=\frac{1}{2 W} \sqrt{W^4-2\left(m_a^2+m_b^2\right) W^2+\left(m_a^2-m_b^2\right)^2}$ denotes the center of mass momentum of  the photon and vector meson, respectively. The differential cross section part of $VMN\to VMN$ process in the formula is given as
\begin{align}\label{eq:xn xn vmd}
\frac{d\sigma_{VMN\to VMN}}{dt}\Bigg|_{t=t_{min}}=\frac{1}{64}\frac{1}{m_{V}^2(\lambda^2-M_{N}^2)}|F_{VMN}|^2
\end{align}
where the nucleon energy is $\lambda=(W^{2}-m^{2}_{V}-M^{2}_{N})/(2m_{V})$ \cite{vmd}, and at the low energy region, the elastic scattering amplitude of $VMN\to VMN$ process is taken as \cite{antip}
\begin{equation}\label{eq:F vmd}
\begin{aligned}
F_{VMN}&\simeq r_{0}^3 d_{2}\frac{2\pi^2}{27}\left(2M_{N}^2-\left\langle N\left|\sum\limits_{h=u,d,s}m_{h}\bar q_{h}q_{h}\right|N\right\rangle\right)
\\
&=r_{0}^3 d_{2}\frac{2\pi^2}{27}2M_{N}^2(1-b)
\end{aligned}
\end{equation}
where the Bohr radius and the Wilson coefficient are \cite{antip,d2}
\begin{align}
&r_{0}= \frac{4}{3\alpha_{s} m_{q} }
\\
&d_{n}  =\left(\frac{32}{N_{c}}\right)^2\sqrt{\pi}\frac{\Gamma(n+5/2)}{\Gamma(n+5)}
\end{align}
respectively. Where $\alpha_{s}$ is the strong coupling constant and $m_{q}$ is the mass of the constituent quark, the relevant parameters are listed in Table  \ref{paremeter}.

As we can see, the above formula relates the vector meson photoproduction cross section to the parameter $b$, besides the trace anomaly can be represented by $\frac{1}{4}(1-b)M$ when away from the chiral limit. Thus the trace anomaly contribution can be studied through VMD model. However, the four-momentum transfers $t_{min}$ varies significantly with $W$ while $t_{thr}=m_{V}^2M_{N}/(m_{V }+M_{N})$ stay steady \cite{igor}. The varies of $t_{min}$ leads to an energy dependence of $\frac{d\sigma_{\gamma N\to VMN}}{dt}\Bigg|_{t=t_{min}}$, ends up resulting the energy dependence of the trace anomaly.
In Reference \cite{thr,igor}, the relationship between the differential and total cross section at the near-threshold is given as
\begin{equation}\label{eq:Fsigma}
\begin{aligned}
\frac{d\sigma_{\gamma N\to VMN}}{dt} \Bigg|_{t=t_{min},W=W_{thr} } 
&=\frac{d\sigma_{\gamma N\to VMN}}{dt} \Bigg|_{t=t_{thr},W=W_{thr}}
\\
&=\frac{\sigma_{\gamma N\to VMN}( W_{thr} )   }{4 |p_{\gamma} |\cdot  |p_{V} | } 
\end{aligned}
 \end{equation}
In order to decrease the energy dependence, we would apply the last two methods to study the trace anomaly contribution.

\begin{table}
\centering
\caption{The relevant parameters of $\rho, \omega$ and $\phi$ \cite{pdg,0.330,Kou 2021,phi 0.770,OR alpha s,machine}.}
\label{paremeter} 
\setlength{\tabcolsep}{2.2mm}{\begin{tabular}{cccccc}
\hline\hline\noalign{\smallskip}
Meson&$\Gamma_{e^+ e^-}$(keV)&$m_q$(GeV)&$m_V$(GeV)&$\alpha_{A}$&$\alpha_{B}$\\
\noalign{\smallskip}\hline\noalign{\smallskip}
$\rho$&$7.04$&$0.330$&$0.770$&$0.439$&$0.601$\\
$\omega$&$0.60$&$0.330$&$0.782$&$0.460$&$0.595$\\
$\phi$&$1.27$&$0.486$&$1.019$&$0.770$&$0.506$\\
\noalign{\smallskip}
\hline\hline
\end{tabular}}
\end{table}

\section{result and discussion}\label{sec:result}
The $\alpha_{A}$ of the three vector mesons come from several previous work, where the $\alpha_{A}$ for $\omega$ and $\rho$ are taken from the estimates based on a relativistic quantum-field model  \cite{OR alpha s}, and for $\phi$, the $\alpha_{A}$ is taken from the calculations based on a quark potential model  \cite{phi 0.770}. $\alpha_{B}$ are all derived from the machine learning prediction \cite{machine}, which varies with the energy scale $Q$. Here the respective meson mass is taken as the corresponding energy scale \cite{meson}.

\begin{figure}[h]
	\centering        
\includegraphics[width=0.48\textwidth]{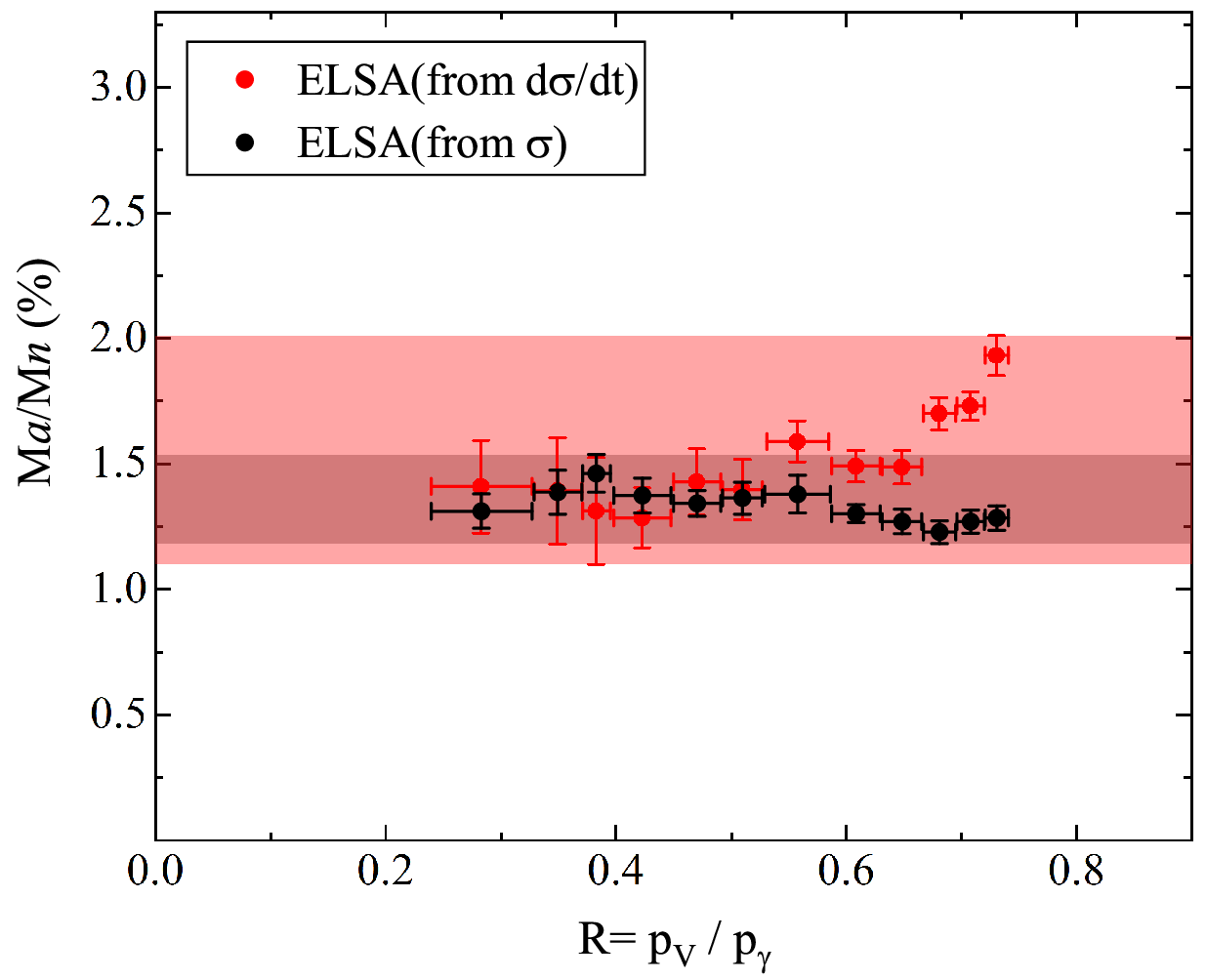}
\caption{The neutron trace anomaly contributions $M_{a}/M_{N}$ as a function of $R$ extracted from the experimental $\omega$ differential and total cross section at $\alpha_{A}$ \cite{OR alpha s,omega data}. The solid red and black circles represent the result from the differential and total cross section, respectively. Bands in the corresponding colour are the error bar.}
  \label{Fig1} 
\end{figure}

After determining the relevant parameters, the calculations are carried out. Firstly, based on the experimental differential and total cross section of $\omega$ at $W\in [1.75,2.15]$ measured by ELSA \cite{omega data}, the neutron trace anomaly contributions were extracted under the framework of VMD model. Results at $\alpha_{A}$ were listed in Table \ref{omega} and shown in Fig. \ref{Fig1} as a function of $R$. The results from the differential and the total section are slightly different due to the numerical difference between the two sections. In order to consider the outcomes in both cases, we process all of the results in the way of root-mean-square, which is $1.43^{+0.58}_{-0.33}\ \%$.

\begin{table}\small
\caption{\label{omega} The neutron trace anomaly contribution extracted from the experimental $\omega$ photoproduction cross section at $\alpha_{A}$ \cite{OR alpha s,omega data}. The average is $1.43^{+0.58}_{-0.33}\ \%$.}
\centering
\begin{ruledtabular}
\centering
\begin{tabular}{cccc}
W (GeV) & $M_{a}/M_{N}$ ($\%$) & W (GeV) & $M_{a}/M_{N}$ ($\%$) \\
 &(from $d\sigma/dt)$& &(from $\sigma$)\\
\hline
$1.76$  & $1.41\pm0.18$ & $1.76$  & $1.31\pm0.07$ \\
$1.78$  & $1.39\pm0.21$ & $1.78$  & $1.39\pm0.09$ \\
$1.79$  & $1.31\pm0.21$ & $1.79$  & $1.46\pm0.07$ \\
$1.81$  & $1.28\pm0.12$ & $1.81$  & $1.37\pm0.07$ \\
$1.84$  & $1.43\pm0.13$ & $1.84$  & $1.34\pm0.05$ \\
$1.86$  & $1.40\pm0.12$ & $1.86$  & $1.36\pm0.06$ \\
$1.90$  & $1.59\pm0.08$ & $1.90$  & $1.38\pm0.08$ \\
$1.95$  & $1.49\pm0.06$ & $1.95$  & $1.30\pm0.04$ \\
$2.00$  & $1.49\pm0.07$ & $2.00$  & $1.27\pm0.05$ \\
$2.04$  & $1.70\pm0.06$ & $2.04$  & $1.23\pm0.06$ \\
$2.09$  & $1.73\pm0.06$ & $2.09$  & $1.27\pm0.05$ \\
$2.13$  & $1.93\pm0.08$ & $2.13$  & $1.28\pm0.05$
\end{tabular}

\end{ruledtabular}
\end{table}

\begin{figure}[h]
	\centering        
	\includegraphics[width=0.48\textwidth]{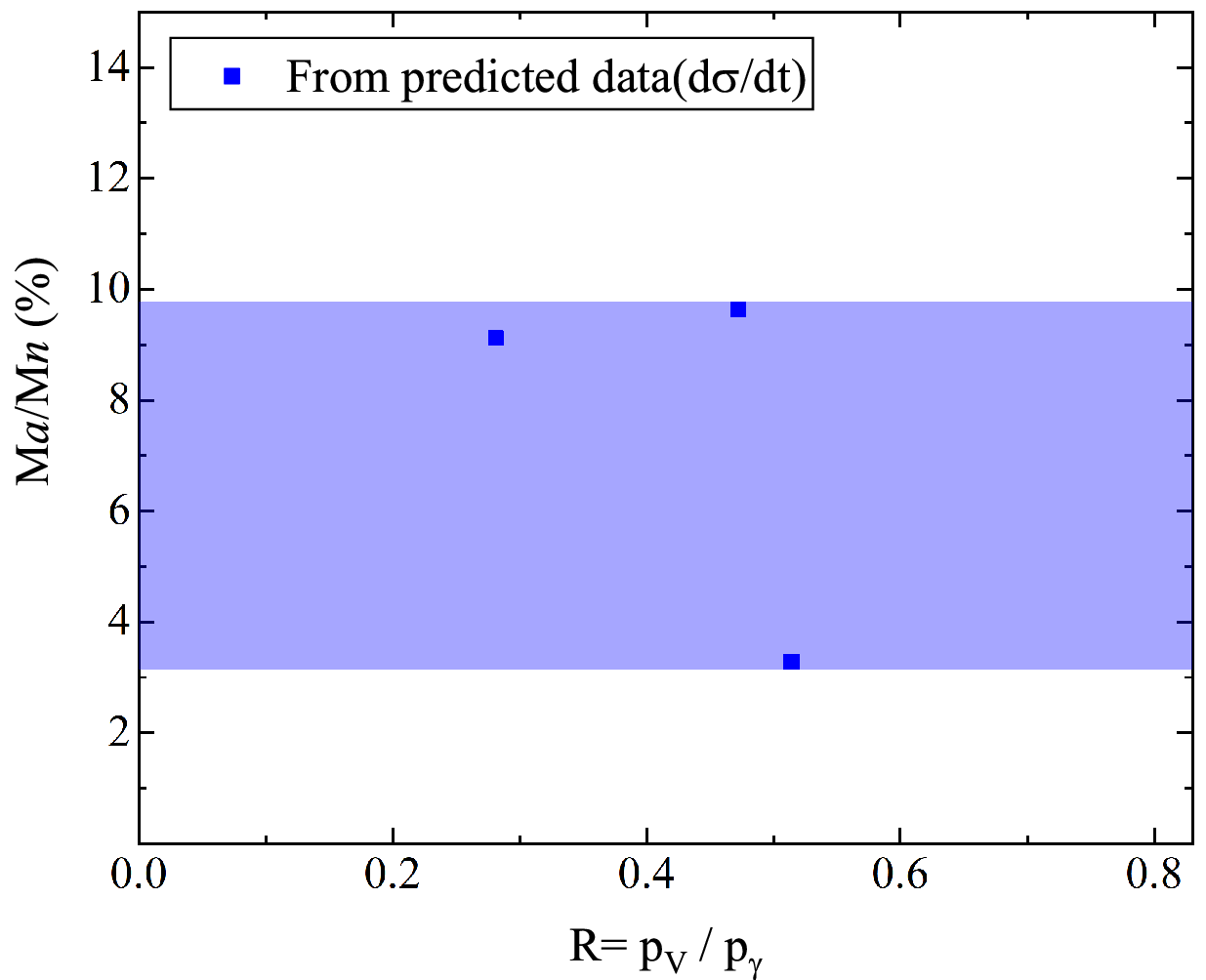}
\caption{The neutron trace anomaly contributions from the predicted $\phi$ differential cross section at $\alpha_{A}$ as a function of $R$ \cite{phi 0.770,phi data}, which is represented by the solid blue square, meanwhile the blue band is the error bar.}
  \label{Fig2} 
\end{figure}

Using the same way of $\omega$, the neutron trace anomaly contributions were extracted from the SLAC differential cross section \cite{rho data} of $\rho$ photoproduction at W=3.87GeV as $1.50\pm0.22$\% under the condition of $\alpha_{A}$. In addition, the neutron trace anomaly contributions also extracted from the predicted $\phi$ differential cross section \cite{phi data} at different energies under $\alpha_{A}$, shown in Table \ref{phi} and Fig. \ref{Fig2} as a function of $R$, the root mean square of the results is $7.89^{+1.74}_{-4.61}$\%. However, the results from $\phi$ photoproduction process vary greatly under different energies, which is caused by the uncertainty of the prediction. So this set of results is for reference only.
                                                            
\begin{table}
\centering
\caption{The neutron trace anomaly contribution extracted from the predicted $\phi$ photoproduction differential cross section at $\alpha_{A}$ \cite{phi 0.770,phi data}. The average is $7.89^{+1.74}_{-4.61}\ \%$.}
\label{phi} 
\setlength{\tabcolsep}{7mm}{\begin{tabular}{cccc}
\hline\hline\noalign{\smallskip}
$W$(GeV)&2.01&2.12&2.15\\
\noalign{\smallskip}\hline\noalign{\smallskip}
$M_a/M_N(\%)$&9.12&9.63&3.27\\
\noalign{\smallskip}
\hline\hline
\end{tabular}}
\end{table}

\begin{table}
\centering
\caption{The trace anomaly contribution RMS of neutron at two sets of $\alpha_{s}$ \cite{phi 0.770,OR alpha s,machine}.}
\label{compare alpha s} 
\setlength{\tabcolsep}{4.5mm}{\begin{tabular}{cccc}
\hline\hline\noalign{\smallskip}
Meson&$\rho$&$\omega$&$\phi$\\
\noalign{\smallskip}\hline\noalign{\smallskip}
$M_a/M_N(\%)$&$1.50\pm0.22$&$1.43^{+0.58}_{-0.33}$&$7.89^{+1.74}_{-4.61}$\\
(from $\alpha_{A}$)& & &\\
$M_a/M_N(\%)$&$3.85\pm0.56$&$3.10^{+1.26}_{-0.72}$&$2.24^{+0.50}_{-1.31}$\\
(from $\alpha_{B}$)& & &\\
\noalign{\smallskip}
\hline\hline
\end{tabular}}
\end{table} 

Under the VMD model, the trace anomaly contribution of the neutron has been extracted on the above, where $\alpha_{s}$ are derived from several different physical models. Another purpose is to discuss the effect of $\alpha_{s}$ variation on the anomaly contribution. We extract the anomaly contribution at $\alpha_{B}$ with the same method through three vector meson photoproduction processes, $\alpha_{B}$ comes from the pure prediction by data learning and has no model dependence \cite{machine}. The trace anomaly contributions RMS of the neutron mass at different $\alpha_s$ are compared in Table \ref{compare alpha s}. Combined with the value of $\alpha_{s}$ shown in Table \ref{paremeter}, one finds that the trace anomaly contribution is very sensitive to it. That indicates $\alpha_{s}$ is a crucial uncertain physical quantity for itself and for calculating the nucleon anomaly contribution.

In works \cite{I} and \cite{light}, the trace anomaly contributions of the proton mass were extracted through the heavy and light vector meson photoproduction processes. We display the comparison of the proton and neutron results at the same set of $\alpha_{s} \ (\alpha_{A})$ in Table \ref{compare p n}, results indicating that the trace anomaly contributions between neutron and proton from the same photoproduction are close considering the error bar, although the results from the different process are still different from each other.

As for the percentage of the nucleon anomaly contribution, although the results obtained with the present calculation accuracy from our works show that it is small, it cannot be taken as the final conclusion yet, the specific reasons will be discussed in the next section.

\begin{table}
\centering
\caption{The trace anomaly contribution RMS of nucleons at $\alpha_{A}$ \cite{light,phi 0.770,OR alpha s}.}
\label{compare p n} 
\setlength{\tabcolsep}{3.5mm}{\begin{tabular}{cccc}
\hline\hline\noalign{\smallskip}
Meson&$\rho$&$\omega$&$\phi$\\
\noalign{\smallskip}\hline\noalign{\smallskip}
$M_a/M_N(\%)$&$0.53\pm0.15$&$1.47 \pm0.48$&$3.63 \pm0.64$\\
(proton)& & &\\
$M_a/M_N(\%)$&$1.50\pm0.22$&$1.43^{+0.58}_{-0.33}$&$7.89^{+1.74}_{-4.61}$\\
(neutron)& & &\\
\noalign{\smallskip}
\hline\hline
\end{tabular}}
\end{table}

\section{SUMMARY}\label{sec:sum}
In this work, we calculate the trace anomaly contribution of the neutron mass through the light vector meson photoproduction processes for the first time based on the VMD model. The numerical results obtained from the three processes are all relatively small, which are $1.43^{+0.58}_{-0.33}\ \%$ from the $\omega$ photoproduction, $1.50\pm0.22$\% from the $\rho$ photoproduction and $7.89^{+1.74}_{-4.61}$\% from the $\phi$ photoproduction.
At the same time, results show that under the same parameter, the trace anomaly contribution of proton and neutron are very close, indicating that the mass distribution and quark-gluon structure inside them may be similar.
And we also study the influence of parameter $\alpha_{s}$ on the nucleon anomaly contribution and proved that $\alpha_{s}$ has a significant affection on it. In addition, the accuracy of the experimental data of photoproduction cross sections and the differences of vector meson production mechanisms also affect the extraction of trace anomaly contribution of nucleon.

In terms of experimental data, the current shortcoming is that the neutron target photoproduction data is insufficient. However, the proton target photoproduction data is relatively abundant, we consider that there is a possibility of using machine learning algorithms to predict the cross section of $J/\psi$ photoproduction off neutron. Besides, the physical model could be effective in predicting the neutron target photoproduction data due to the exclusive reaction channels of the $J/\psi$ production. We will explore the above aspects in our follow-up work. At the same time, more high-precision experimental measurements for the photoproduction are expected.

As for the differences in vector meson production mechanisms, we found that the difference in mesons’ related properties visibly affects the nucleon trace anomaly contribution by analyzing the results from different photoproduction processes under the VMD model. For example, the decay width of mesons has a significant effect on the proton anomaly contribution. The $\omega$ meson and the $\rho$ meson have a similar mass, corresponding $\alpha_{s}$, and component quark mass, but there is a large gap between the anomaly contribution results calculated from the two processes, which comes from the large decay width of $\rho$. These results bring the anomaly contribution of proton mass a notable error, if the partial width $\Gamma_{\rho \to e^+ e^-}$ of $\rho$ is set as $\Gamma_{\omega \to e^+ e^-}$, the results of the proton anomaly contribution will be very close \cite{light}. And based on the verification in Sec. \ref{sec:result}, we have learned that $\alpha_{s}$ is an important parameter noticeably affecting the nucleon trace anomaly contribution. At the same time, $\alpha_{s}$ also has a large uncertainty in the low energy range, this is due to the complexity of the non-perturbative QCD. Meanwhile, the corresponding scales of $\alpha_{s}$, which are taken as the meson mass, are included within this range exactly. The appreciable error of $\alpha_{s}$ and its significant effect on the nucleon trace anomaly contribution leads to the uncertainty of the anomaly contribution. We will follow on using the machine learning algorithms to study $\alpha_{s}$, to explore the physics involved and try to correct its accuracy. In fact, $\alpha_{s}$ has great accuracy both experimentally and theoretically in the high $Q$ range. If photoproduction experimental data of heavier mesons are available, the uncertainty from $\alpha_{s}$ will be eliminated through the calculation from these processes.

The scattering length of vector meson-nucleon(VMN) can also be calculated under the VMD model, and the related studies are abundant \cite{igor,thr,41,42,43,44}. Meanwhile, the uncertainty of the involved parameters in the calculation is small. Therefore, we will consider exploring the possibility of associating the VMN scattering length with the nucleon trace anomaly contribution under the VMD model, trying to find another way to reveal the nucleon trace anomaly contribution.

In a word, although the current numerical results of the nucleon trace anomaly contribution are small, this can only be considered a partial conclusion due to a series of uncertainties. We will conduct our follow-up based on the above points, aiming to extract a more accurate nucleon trace anomaly contribution.

\begin{acknowledgments}
This work is supported by the National Natural Science Foundation of China under Grants No. 12065014 and No. 12247101, and by the Natural Science Foundation of Gansu province under Grant No. 22JR5RA266. We acknowledge the West Light Foundation of The Chinese Academy of Sciences, Grant No. 21JR7RA201.
\end{acknowledgments}

\end{document}